\documentclass[11pt, letterpaper, twoside]{article}
\usepackage[utf8]{inputenc}
\usepackage{mathtools}
\usepackage{nccmath}
\usepackage{amsmath}
\usepackage{xcolor}
\usepackage{authblk}
\usepackage{caption}

\usepackage{comment}

\title{Hellinger's distance to normal distribution as market invariant}
\author[1]{Mesrop T. Mesropyan}
\author[2]{Vardan G. Bardakhchyan}
\affil[1]{Yerevan State University, Chair of Actuarial and Financial Mathmetics, Graduate student}
\affil[2]{Yerevan State Uinversity,  Chair of Actuarial and Financial Mathematics, PhD, lecturer}

\date{mesropyanmesrop@list.ru; vardan.bardakchyan@ysu.am}

\begin{document}
\maketitle

\begin{abstract}
    Main purpose of distance based portfolio constructions is in portfolio imitation. Here we construct portfolio based on Hellinger's distance from normal distribution. We empirically found that minimum of this distance drastically varies from market to market. Thus we suppose that it may be regarded as a form of market invariant, in a sense of helpful tool for market segmentation. We analyze its sensitivity. Though mean sensitivity was small it showed extreme sensitivity in some cases.
\end{abstract}


\section{Introduction and motivation}

The hunt of market invariants has motivated authors to incorporate various interdisciplinary techniques.
All of them use slightly different definitions of market invariant, use different techniques to derive them, and for different purposes.

Invariants in general are quantities which are not affected when some transformation of given class occur.
However in financial data they are perceived slightly differently. 
Some authors define market invariants to be random variables or random vectors the distribution of which does not change over time.
Some authors treat market invariants as variables which are enough robust either remaining in small interval when time changes, or experiencing small or no change when data is transformed (e.g. scaled). 

Also there are several purposes for determining market invariant. For example \cite{Entropy} used market invariants to determine market manipulation. Some authors used market invariants as a tool of market segmentation by their characteristics, i.e different values of market invariants corresponded to different segments of market.

Some authors define market invariants as universal constants. 
Latest theory proposed by \cite{Trading 1} take axiomatic approach to define one such possible constant, which later proved good enough empirically \cite{Trading 2, Trading 3}.
They proposed $\frac{R}{N^{3/2}}$, with R total risk traded, and $N$ number of daily bets (or bettors).
This universal constant was applicable to a market in whole.

\noindent Other authors use more intuitive approach to finding invariants.

\cite{Entropy} proposed as market invariant entropy. Their motivation was based on dynamical systems theory. 
Though they name two other possible measure used to identify (describe) non-linear system, namely Lyapunov exponent and correlation dimension.

\cite{String} analyzed foreign exchange markets (namely EUR/USD) series by means of string theory. Their inspiration was from later developments in physics. As invariant they proposed weighted sum of one lag correlations between returns in given time interval with weights following Gibbs distribution.

In these last two papers the market invariant is invariant of specific time series.

On the other hand \cite{Correlations} considered correlations between log-returns of different groups of assets. After subtracting mean log-return, the correlation between new series were checked for invariance. (Four other form of correlations were analyzed and compared).

Our purpose of defining possible market invariant is pure for diagnostic use and differentiating between several markets.

Here we treat market invariant as some measure specific to given market (given class of financial securities) that is enough robust to changes in data. 
Our analysis is based on portfolio construction, and as such techniques described here cannot be applied to one security time series. 

Our analysis is based upon the consideration that log-returns are indeed invariants, most of authors take this as granted (see for example \cite{Meucci}). So we do not perform time series analysis or any other kind of modelling, and treat the log-returns of one time-series as i.i.d r.v.-s.

The main idea is to construct portfolio closest in distribution to Gaussian.
So we do make use of statistical distances. For each level of expected return we find the minimum distance portfolio (expressed in portfolio weights). And among the collection of these "best" portfolios we determine one which has minimum of all distances. The distance found will be the measure (we believe to be a robust), differing from market to market.
Here we use the Hellinger's distance, though other statistical distances can also be used. 
\footnote{In some sense the invariant estimated shows non-gaussianity of the market considered. Thus it is alike Negentropy (\cite{Independent_comp}), but with crucial differences that we hold constant only expectation and we take distance of linear combination a strictly bounded set of random variables. We don't use ExKurtosis measure as well due to the drawbacks described in \cite{Independent_comp}}

The use of normal distribution is inspired by CLT, and thus by the hope that big markets will eventually yield these distances to be 0. The second reason to use normal distribution is based on pure practical considerations. The class of elliptic distributions are easier to analyze, as two parameters (mean and variance) is enough to describe portfolio and it is easier to make any form of predictions.

Market invariant based on portfolio construction means that we analyze properties of linear combinations (in our case convex combinations, with non-negative weights), instead of using any other functional form of given data. One reason for this is obvious: generally trading is linear combination of securities. If something is internally constant in linear combination of given securities returns, it will be useful to derive it explicitly. 
The other reason is that constructing portfolio in the manner is pure static procedure (though based on historical data, when we analyze distributions), so if something remains close to constant, it wouldn't vary much when time passes. So internally formulation in form of static optimization is preferred without regard to specific future time horizons. While most authors take invariant to be internal characteristic of one or several time series over specific time intervals, we prefer to analyze one time characteristics.

We conduct sensitivity analysis on pure empirical basis, to check robustness of the distance. 
The techniques used can be treated as solving portfolio optimization problem, where squared Hellinger's distance is used as risk measure. As thus sensitivity analysis of minimal squared Hellinger's distance is in fact portfolio sensitivity analysis (like ones for mean-variance or VaR measures \cite{VaR, CVaR}).
As Hellinger's distance is formulated by probability measures (distribution functions), the matrix perturbation techniques used by \cite{BestGrauer} cannot be applied.

The paper is organized as follows. In section 2 the general problem of constructing portfolio with minimal Hellinger distance from normal distribution is stated,  computational techniques are showed and result are briefly summarized. In section result of sensitivity analysis are shown. 
The paper ends with conclusion.

\section{Hellinger's distance based portfolio construction}
 
 Classical formulation of mean-variance portfolio is the following.

\begin{equation}
\centering
\left\{\begin{split}
& E(X) = e \\
& Var(X) \rightarrow min \\ 
& w_i \geq 0 \\
& \sum{w_i} =1
\end{split}\right.
 \end{equation}

Solving the system above one get the portfolio with minimum variance for each level of expected return.
 $X_1,\ X_2,\ X_3,\ldots$ represent returns of each asset. 
 We confine ourselves to the case of logarithmic return.
  
$w_i$ represents the percentage (weight) of money in each asset. We assume only long positions i.e. $w_i\geq0$, $\sum w_i=1$.

$$E\left(X\right)=E\left(\sum_{i=1}^{n}{w_i X_i}\right)=\sum_{i=1}^{n}{E\left(w_i X_i\right)=\sum_{i=1}^{n}{w_iE\left(X_i\right)}} $$

Our goal in constructing minimum Hellinger's distance (or squared one) can be formulated by following problem.
\\
\begin{equation}
\centering
\left\{\begin{split}
& E(X) = e \\
& H^{2}(X,N) \rightarrow \min  
\end{split} \right.
\end{equation}
where $H\left(X,N\right)$ - 

In other words Hellinger’s distance which should be the least from normally distributed random variable. Parameters of desired normal distribution are got from solving (1) problem. 
So More generally we solve the following.

\begin{equation}
\centering
\left\{\begin{split}
& E(X) = e \\
& H^{2}\left(X, N\left(e, Var(X^{*}(e))\right)\right) \rightarrow min \\ 
& w_i \geq 0 \\
& \sum{w_i} =1
\end{split} \right.
\end{equation}

with $X^{*}(e)$ being solution of (1).
After solving (3) we take minimum of all Hellinger distances $\min_{e} [H^{2}]^{*}$

We note that the above measure is differs significantly from market to market.

For continuous case

\begin{equation*}
\begin{split}
H^2\left(f,g\right) & =\frac{1}{2}\int{\left(\sqrt{f\left(x\right)}-\sqrt{g\left(x\right)}\right)^2dx} \\
& =1-\int\sqrt{f\left(x\right)g\left(x\right)}dx
\end{split}
\end{equation*}

To solve the problem with discrete data we do binning process\footnote{we exactly use binning, instead of kernel methods as of simplicity of calculations, and one more reason. The kernel methods have one more parameter which (as we checked) will add sensitivity to given solutions. Exactly the method is sensitive to what kernel is used}, which will yield by to continuous distribution represented by simple function. \\
The distance between these kind of distributions will take the following form:
\begin{equation}
\begin{split}
    H^2\left(X,Y\right) & = 
    1-\int_{a}^{b}{\sqrt{fg}dx} \\ & =1-\left(\int_{a}^{a_1}{+\int_{a_1}^{a_2}{+\ldots+\int_{a_{n-1}}^{b}{\sqrt{fg}dx}}}\right) \\
    &=1-\sum_{i} O_i\int_{a_i}^{a_{i+1}}\sqrt{g(x)}dx
\end{split}
\end{equation}

with $f\left(x\right)={O_i}^2$ if $x\in[a_i,a_{i+1}]$ (with $a=a_0$ and $a_n=b$). Generally $O_i$-s are any numbers, not bound to be different. We take the interval cut enough fine to have one value for each interval. \\
Whenever the counterpart distribution is normal (i.e. g(x) is density of normal distribution), we will the following:
\begin{equation}
\begin{split}
\int_{c}^{d} & {\sqrt{g(x)}dx}  =\left(\frac{1}{\sigma\sqrt{2\pi}}\int_{c}^{d}{e^{-\frac{\left(x-\mu\right)^2}{2\sigma^2}}dx}\right)^\frac{1}{2} \\ & =\frac{1}{\sigma^\frac{1}{2}\left(2\pi\right)^\frac{1}{4}}\int_{c}^{d}{e^{-\frac{\left(x-\mu\right)^2}{4\sigma^2}}dx} \\
& =\frac{1}{\sigma^\frac{1}{2}\left(2\pi\right)^\frac{1}{4}}\int_{c}^{d}{e^{-\frac{\left(x-\mu\right)^2}{4\left(\sigma\sqrt2\right)^2}}dx} \\
& 
\overset{\sigma_1 = \sigma \sqrt{2}}{\longrightarrow} \frac{1}{\sigma^\frac{1}{2}\left(2\pi\right)^\frac{1}{4}}\frac{1}{\sigma_1\sqrt{2\pi}}\sigma_1\sqrt{2\pi}\int{e^{-\frac{\left(x-a\right)^2}{2\sigma_1^2}}dx} \\
& =\sqrt2\sigma^\frac{1}{2}\left(2\pi\right)^\frac{1}{4}\left(F_N\left(d|\mu;2\sigma^2\right)-F_N\left(c|\mu;\ 2\sigma^2\right)\right)
\end{split}
\end{equation}

\vskip 4mm

We noted that for different types of market the minimum squared Hellinger's distance is different.

 More formally we found the following result for 3 different markets - Stock market of specific types of stock,
 exchange market and purely simulated market of correlated student's distributions with different degrees of freedom.

We got the following result \footnote{We used 4 randomly generated student distribution's with 4, 3, 3 and 2 degrees of freedom. We got correlated variables with Cholesky decomposition of randomly taken correlation matrix. The data was taken to be 810 data, for each (approx 3 years).  We took 4 stocks of nearly the same branch (Cisco, Intel, Microsoft, NVIDIA), for the period of 3 years. And lastly we analyzed foreign exchange market data. We took data up to 7th month of 2020, of approximately 3 years (After that war started at Nagorno-Karabagh, which had influence to our money market, and exchange rates.)} \footnote{The data is in open access, so anybody can conduct the same analysis.}

\begin{center}
\captionof{table}{Squared Hellinger's distances calculated for 3 different markets}
    \begin{tabular}{||c|c|c|c||}
        \hline
         & Simulated data market & Stock Market & Exchange market data \\
         \hline
        $H^2$ & 0.0127 & 0.027 & 0.0149 \\
        \hline
    \end{tabular}
\end{center}

\section{Sensitivity Analysis}

Generally under sensitivity analysis in optimization problems it is meant sensitivity checking depending of initial conditions or parameters. In this general form sensitivity of our problem seems infeasible as we deal with general distributions. So we checked sensitivity by randomly "editing" data. 
 
\noindent We did 2 types of sensitivity check. 
 
First we changed randomly taken portion of data, by random but restricted magnitude. We then computed percentage change, per 5\% \footnote{smaller choices make to minor differences} of changed data. Here we report both absolute values of change and percentage. 
\noindent Next we checked effect of changes in binning number.\footnote{All the analysis were done by Python 3.10. We used Excel 2019 (for data preprocessing)}

For the last part it is reasonable to expect that adding new variable will decrease the minimal Hellinger distance by CLT. However when we speak about initial small number of variables it is not straightforward with adding one new.

For binning procedure change of number of bins must be expected to bring to extreme changes in Hellinger. 

\noindent For the first two procedures nothing exact can be expected.

\noindent We got the following results \footnote{We randomly changed a portion of initial data by random amount up to 1000 times and calculated average change in each case}

\begin{center}
\captionof{table}{Change in squared Hellinger due to changing up to 5\% of data of maximum of 5\% of their value}
    \begin{tabular}{||c|c|c|c||}
        \hline 
          & Simulated data & Stock & ForEx \\
         \hline
        Percentages & 3.02\% & 1.01\% & 5.51\% \\
        \hline
        Absolute value & 0.00039125 & 0.000299 & 0.00092 \\
        \hline
    \end{tabular}
\end{center}

We got the following results for binning and for adding new variables. 

\begin{center} 
\captionof{table}{Change in Hellinger due to bin number change by 1}
    \begin{tabular}{||c|c|c|c||}
        \hline 
         & Simulated data market \\
         \hline
        Binning & $\sim 10\% $ \\
        \hline
    \end{tabular}
\end{center}

\section{Discussion and conclusion}

We have conducted a general sensistivity analysis for the minimum squared Hellinger's distance for 3 different financial markets. 

There are several reasons to cast doubt about the results presented.

These sensitivity analysis lacks statistical background, i.e. we can't check statistically whether the results found reveal the true picture. However we have conducted big enough number of simulation to check.

Second, the optimization procedure is not smooth one. Thus the initial guess of weights had crucial role. (We checked for enough fine grid, but we didn't combine finite difference techniques). 

But generally the procedure revealed that only substantial changes can bring to to changes minimum squared Hellinger's distance. 
Though we found generally less that 5\% change in responce to 5\% change of initial data, one should note that we considered 5\% change in only 5\% of data. Anyway, small changes had extremely less effect. 

Obviously adding some dominant (in mean-variance sense) and normally distributed random variables to the market, it would drug the Hellinger down. So the measure proposed is showing how you efficient frontier is composed of close to normal elements. 

Two things to note, is that obviously if all returns in market considered are normally distributed, one will get 0 Hellinger. 
However, though it can be supposed, that the bigger the market, the smaller Hellinger it would have, there are some source of doubt about it. 
As we consider the distance to normal with mean and variance that of efficient frontier. As it is generally known, the Markowitz's mean-variance portfolio tend to use only small number of instrument for each level mean return. 
If this small number can be combined to get closer to normal, the Hellinger will decrease substantially, otherwise the changes will be not that big.

Much further work should be done to find out whether the minimal Hellinger is indeed market invariants or not.

\end{document}